\documentclass[prl,twocolumn,showpacs,superscriptaddress,floatfix,longbibliography]{revtex4-1}
\usepackage{graphicx,amsfonts,amssymb,amsmath,xspace,dsfont}
\usepackage[colorlinks=true,citecolor=green,linkcolor=red,urlcolor=blue]{hyperref}

\newlength{\ldag}
\newcommand{\bn}{{\boldsymbol{n}}}

\newcommand{\bk}{{\boldsymbol{k}}}

\begin{document}

\title{Low-energy effective theory of the toric code model in a parallel field}

\author{Julien Vidal}
\email{vidal@lptmc.jussieu.fr}
\affiliation{Laboratoire de Physique Th\'eorique de la Mati\`ere Condens\'ee,
CNRS UMR 7600, Universit\'e Pierre et Marie Curie, 4 Place Jussieu, 75252
Paris Cedex 05, France}

\author{S\'ebastien Dusuel}
\email{sdusuel@gmail.com}
\affiliation{Lyc\'ee Saint-Louis, 44 Boulevard Saint-Michel, 75006 Paris, France}

\author{Kai Phillip Schmidt}
\email{schmidt@fkt.physik.uni-dortmund.de}
\affiliation{Lehrstuhl f\"ur theoretische Physik, Otto-Hahn-Stra\ss e 4, D-44221 Dortmund, Germany}

%------------------------------------------------------------------------------

\begin{abstract}
 We determine  analytically the phase diagram of the toric code model in a parallel magnetic field
 which displays three distinct regions. Our study relies on two high-order
 perturbative expansions in the strong- and weak-field limit, as well as a
 large-spin analysis.  Calculations in the topological phase establish a quasiparticle 
 picture for the anyonic excitations.
 We obtain two second-order transition lines that merge with a
 first-order line giving rise to a multicritical point as recently suggested by
 numerical simulations. We compute the values of the corresponding critical
 fields and exponents that drive the closure of the gap. We also give the one-particle dispersions of the anyonic quasiparticles inside the topological phase.
\end{abstract}


\maketitle

%
%
%%%%%%%%%%%%%%%%%%%%%%%
%%%%%%%%%%%%%%%%%%%%%%%
\section{Introduction}
%%%%%%%%%%%%%%%%%%%%%%%
%%%%%%%%%%%%%%%%%%%%%%%
%
%
For more than 30 years, lattice gauge theories have been the subject of
intense researches especially in high-energy physics where they aim at
describing quark confinement \cite{Wilson74}. Such theories are deeply related
to topological phase transitions characterized by the absence of local order
parameters \cite{Wegner71,Kogut79}.
One of the most famous models where such transitions occur is the $\mathbb{Z}_2$
gauge and matter theory whose phase diagram in the three-dimensional case (2+1)
has been widely studied by means of various methods
\cite{Fradkin79,Jongeward80,Creutz80,Banks81,Brezin82,Nussinov05_2}.
Interestingly and very recently, this model has been shown to be equivalent to
the toric code model (TCM) in a magnetic field in which one introduces
ancillary (matter) fields together with a gauge-invariance constraint
\cite{Tupitsyn08}.

The TCM was introduced by Kitaev to perform topological
quantum computation \cite{Kitaev03}. This spin model can be solved exactly
and exhibits two kinds of dispersionless excitations, called charges and fluxes,
which have mutual anyonic statistics although each of them are bosons.
In the absence of an external magnetic field, these anyons are localized
on the vertices (charges) and on the plaquettes (fluxes) of a square
lattice. Let us emphasize that the detection of anyonic statistics in the TCM
has been the subject of several experimental proposals in optical lattices
\cite{Jiang08,Aguado08}, although there the TCM appears as a low-energy
effective theory of Kitaev's honeycomb model \cite{Kitaev06,Schmidt08}.

The aim of this paper is to study the influence of a magnetic
field in the TCM. Contrary to a recent study \cite{Tupitsyn08}, we directly
consider the quantum problem instead of using its classical counterpart.
As we will see, the magnetic field gives rise to a nontrivial phase diagram
which displays first-order and second-order transition lines merging in a
topological quantum multicritical point located at the confluent of topological
and ordered phases. Additionally, we provide a quasiparticle (QP)
description of the anyonic excitations in the topological phase.

To investigate this issue, we use several perturbative treatments. First,
we perform a standard (linear) spin-wave analysis \cite{Manousakis91} which
captures quantum fluctuations around the classical ground state and is thus
certainly valid (qualitatively) in the large-field limit.
Second, we compute the perturbative
expansion of the ground-state energy and the gap in the small-field limit and in the
large-field limit by means of the continuous unitary transformations method
\cite{Wegner94,Stein97,Knetter00,Knetter03_1}. These three approaches allow us to propose a consistent picture of the phase diagram. Finally, we also give the dispersion relation in the 1-QP sector in the topological phase.

%
%
%%%%%%%%%%%%%%%%%%%%%%%
%%%%%%%%%%%%%%%%%%%%%%%
\section{Model}
\begin{figure}[t]
 \includegraphics[width=0.55\columnwidth]{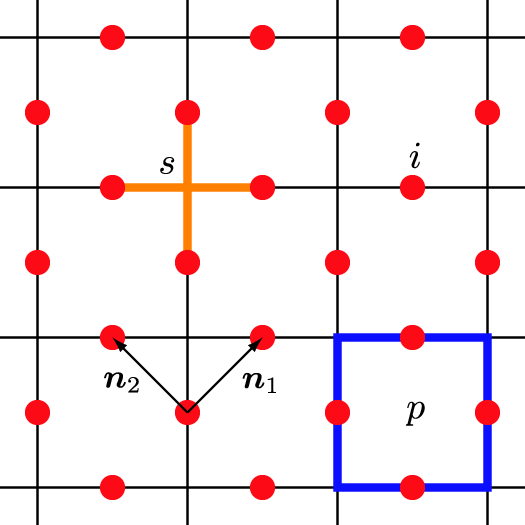}
 \caption{(Color online). A piece of the square lattice on which the TCM is
   defined. Spins (dots) are located on the bond and interact if they share
   either a vertex $(s)$ or a plaquette $(p)$.}
 \label{fig:lattice}
\end{figure}

%%%%%%%%%%%%%%%%%%%%%%%
%%%%%%%%%%%%%%%%%%%%%%%
%
%
We consider the following Hamiltonian
% 
%
%%%%%%%%%%%%%%
\begin{equation}
 \label{eq:ham}
 H = - J_s \sum_{s} A_s  - J_p \sum_{p} B_p
 - h_x \sum_i \sigma_i^x - h_z \sum_i \sigma_i^z, 
\end{equation} 
%%%%%%%%%%%%%%
%
%
where the $\sigma_i^\alpha$'s are the Pauli matrices,
$A_s=\prod_{i \in s} \sigma_i^x$, and $B_p =\prod_{i \in p} \sigma_i^z$.
Subscripts $s$ and $p$ refer respectively to sites and plaquettes of a square
lattice whereas $i$ runs over all bonds where spin degrees of freedom are
located (see Fig.~\ref{fig:lattice}).

Up to a global normalization, the parameter space of Hamiltonian
(\ref{eq:ham}) is three-dimensional.
Here, we focus on the two-dimensional subspace defined by $J_s=J_p=J$ which, for
$h_x=h_z=0$, coincides with the TCM \cite{Kitaev03}.
We emphasize that this subspace is not the same as the one considered in
Ref.~\cite{Tupitsyn08} where the variables ($J_s, J_p, h_x, h_z)$ are linked
via the mapping onto the isotropic $\mathbb{Z}_2$ gauge Higgs model.
Consequently, {\em one cannot compare our results} with the numerical data
\cite{Tupitsyn08} but, as we shall see, our phase diagram displays similar
qualitative features.

Let us mention that the single-component magnetic field case has also been
addressed recently \cite{Trebst07,Hamma08} but its low-energy properties are exactly the same as those of the celebrated two-dimensional Ising model in a transverse field whose phase diagram
has been determined accurately many years ago \cite{He90}.

%
%
%%%%%%%%%%%%%%
%%%%%%%%%%%%%%
%
%

%
%
%%%%%%%%%%%%%%%%%%%%%%%
%%%%%%%%%%%%%%%%%%%%%%%
\section{Linear spin-wave theory}
%%%%%%%%%%%%%%%%%%%%%%%
%%%%%%%%%%%%%%%%%%%%%%%
%
%
As a first approach, let us compute the low-energy spectrum of $H$ in the
semi-classical (large-spin) limit. Therefore, we \mbox{consider} the Hamiltonian $H_S$
obtained by replacing the Pauli matrices by SU(2) spin-$S$ operators ($\sigma_i^\alpha \rightarrow S_i^\alpha/S$) which are considered as classical vectors in the large-$S$ limit [${\bf S_i}=S (\sin \theta_i \cos \phi_i, \sin \theta_i \sin \phi_i, \cos \theta_i)$]. 
Assuming that the ground state is given by a uniform configuration
($\theta_i=\theta, \phi_i=0, \, \forall i$) \cite{Tempo}, we are led to minimize
the corresponding classical energy $E$. 
Next, we use the Holstein-Primakoff representation of the spin operators to compute the leading $1/S$ corrections and we obtain the excitation energies 
$\Delta_\bk=\frac{1}{S}\sqrt{\varepsilon_\bk^2- \gamma_\bk^2}$, where
%
%
%%%%%%%%%%%%%%%%%%%%
\begin{equation}
 \varepsilon_\bk = \gamma_\bk
 + 2 J \big( \sin^4\theta_0 + \cos^4 \theta_0 \big)
 + h_x \sin \theta_0 + h_z \cos \theta_0, 
  \label{eq:eps_k}
\end{equation}
and
\begin{eqnarray}
 \gamma_\bk &=& - J \cos^2 \theta_0 \sin^2 \theta_0
 \Big\{ 2 \cos(\bk.\bn_1)+2 \cos(\bk.\bn_2)\nonumber \\
 && +\cos\big[\bk.(\bn_1+\bn_2)\big]+\cos\big[\bk.(\bn_1-\bn_2)\big] \Big\}.
 \quad \quad \quad
\end{eqnarray}
%%%%%%%%%%%%%%%%%%%%
%
Here, $\bn_1$ and $\bn_2$ are unit vectors displayed in Fig.~\ref{fig:lattice} and $\theta_0$ is the angle which minimizes $E$.

Within this approach we find that the low-energy spectrum is gapped for all
values of the field except for $h_x=h_z=\sqrt{2} J=h_{\mathrm c}^\mathrm{sw}$
where it vanishes.
Further, we find that below this value, on the isotropic line $h_x=h_z$,
the classical energy has two degenerate minima with the same
characteristics. 
This leads us to conclude that the segment $0<h_x=h_z<h_{\mathrm c}^\mathrm{sw}$ is a first-order transition line which ends
at the second-order critical point $h_{\mathrm c}^\mathrm{sw}$. Note that the value of $h_{\mathrm c}^\mathrm{sw}$, being independent
of $S$, certainly differs from the actual value for $S=1/2$.

Although one can reasonably believe in the qualitative features of the spin-wave scenario in the large-field limit, it certainly fails for small fields. Indeed, in this limit, the ground state is
far from being a separable state for $S=1/2$ as can be inferred from the
zero-field case. In addition, for $S>1/2$, $H_S$ does not commute with $A$'s and $B$'s in zero field which is a key ingredient of the TCM's topological character.

%
%
%%%%%%%%%%%%%%%%%%%%%%%
%%%%%%%%%%%%%%%%%%%%%%%
\section{Large-field limit $(h_x,h_z \gg J)$}
%%%%%%%%%%%%%%%%%%%%%%%
%%%%%%%%%%%%%%%%%%%%%%%
%
%
To determine the value $h_{\mathrm c}$ of the critical field for $S=1/2$, let us investigate the low-energy spectrum of Hamiltonian (\ref{eq:ham}) in the
strong-field limit and on the isotropic line $h_x=h_z=h$.
For $J=0$, the ground state of $H$ is fully polarized in the field direction, and
elementary excitations are static single spin-flips with energy cost
$2^{3/2}h$. For $J > 0$ and setting $h=2^{-3/2}$, Hamiltonian (\ref{eq:ham}) can be recast into
%
%
%%%%%%%%%%%%%%%%%%%%
\begin{equation}
 \label{eq:ham_SF}
 H=-N+Q+\sum_{n=0,\pm 1,\pm 2,\pm 3,\pm 4}T_n,
\end{equation}
%%%%%%%%%%%%%%%%%%%%
%
%
where the operators $T_n$ are proportional to $J$ and change the number of
excitations $Q$ by $n$, {\em i.e.}, $[Q,T_n]=n\, T_n$. Their expressions are easily
obtained but are too lengthy to be given here.
To study this Hamiltonian, we used the perturbative continuous unitary
transformations (PCUT) method \cite{Stein97,Knetter00,Knetter03_1},
which allows one to construct, order by order (in $J$), an effective
Hamiltonian which is unitarily equivalent to $H$ but conserves the number of
QPs. The ground state of the effective Hamiltonian is the
0-QP state, whereas the lowest-excited states lie in the
1-QP sector.
We thus have access to the ground-state energy per spin $e_0$, to the
dispersion of the QP and consequently, to the gap $\Delta$. 
At order $5$ in $J$, one has
%
%
%%%%%%%%%%%%%%%%%%%%
\begin{eqnarray}
 \label{eq:gseLF}
 e_0&=& -\frac{1}{2}-\frac{J}{4}-\frac{79}{192} J^2+\frac{251}{1152} J^3
 -\frac{4859243}{15482880} J^4\nonumber\\
 &&+\frac{1503945223}{3251404800}J^5, \quad \\
 \label{eq:gapLF}  
 \Delta&=& 1-J-\frac{11}{48}J^2+\frac{71}{256}J^3 -\frac{1101497}{552960} J^4
 \nonumber\\
 &&+\frac{13570006967}{1300561920}J^5.
\end{eqnarray}
%%%%%%%%%%%%%%%%%%%%
%
%

A standard Dlog Pad\'e approximants analysis for the gap leads to
$h_{\mathrm c}=0.48(2)J$ which, as anticipated, strongly differs from the
spin-wave value ($h_{\mathrm c}^\mathrm{sw}=\sqrt{2} J$) but confirms the
existence of a critical point on the isotropic line and undoubtedly provides a
more accurate determination of its location.
Moreover, in the vicinity of this point and on this line, a $[2,2]$-Dlog Pad\'e approximant yields
$\Delta \sim (h-h_{\mathrm c})^\nu$ with $\nu\simeq 0.73$ instead of $\nu=1/2$ in the linear spin-wave approach. 
This result is hardly compatible with an Ising-type critical point suggested in the
$\mathbb{Z}_2$ gauge Higgs model \cite{Brezin82} for which  $\nu=0.6301(8)$ \cite{Bloete95}, but this is clearly due to the relatively low order of our expansion. 
%(\ref{eq:gapLF}). 

%
%
%%%%%%%%%%%%%%
\begin{figure}[t]
 \includegraphics[width=0.7\columnwidth]{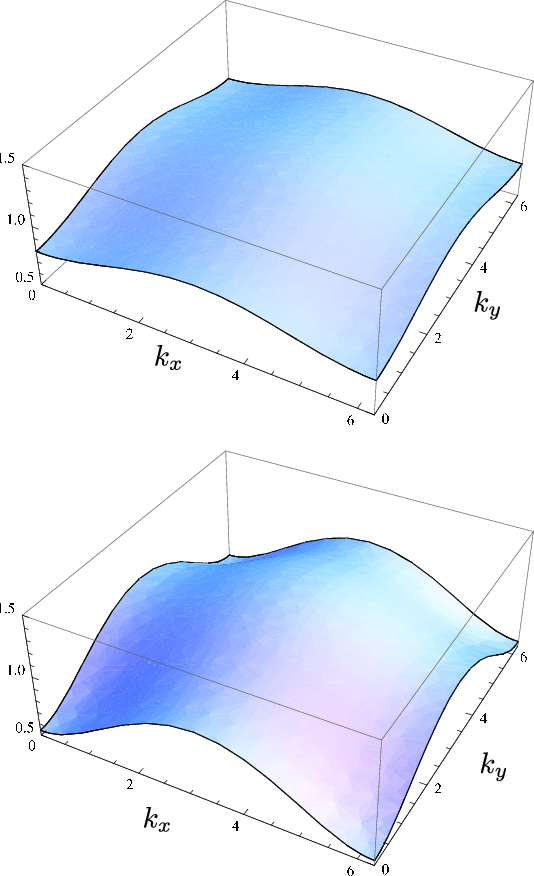}
 \caption{(Color online). One-flux (up) and one-charge (down) dispersions  for $h_z=0.1$, $h_x=0.05$, and $J=1/2$ in the reciprocal lattice. In this case ($h_x< h_z$), the gap is given by the minimum of the one-charge dispersion  [$\Delta \simeq 0.54$ as can be checked from Eq.~(\ref{eq:gapSF})].}
 \label{fig:Disp}
\end{figure}
%%%%%%%%%%%%%%
%
%

%
%
%%%%%%%%%%%%%%%%%%%%%%%
%%%%%%%%%%%%%%%%%%%%%%%
\section{Small-field limit $(h_x,h_z \ll J)$}
%%%%%%%%%%%%%%%%%%%%%%%
%%%%%%%%%%%%%%%%%%%%%%%
%
%
This region is, by far, the most interesting and challenging one. As explained above, 
for $h_x=h_z=0$, Hamiltonian (\ref{eq:ham}) is exactly the TCM and one thus expects the
system to be in a topological phase, whose breakdown should occur at finite
fields only. The ground state of the TCM has all $A_s=B_p=1$. Elementary
excitations are such that one charge or one flux is present, namely one $A_s=-1$
or one $B_p=-1$, and have an energy cost of $2J$. Note that here, we are assuming open boundary conditions and the
thermodynamic limit which allows one to consider a single (flux or charge)
excitation. On a torus, such configurations are prohibited and one must analyze
states with pairs of fluxes or of charges.

In the TCM, fluxes and charges are static but,  for finite fields,  these excitations acquire some dynamics and can be considered as true QP.
Of course, setting up such a description in this (liquid) topological phase is a highly nontrivial task. 
In the present case, this is made possible thanks to the strict locality of the anyonic excitations for
$h_x=h_z=0$.  Thereafter, we establish  perturbatively this QP picture and we present results for the one-flux and one-charge dispersions in the topological phase. 
The study of the one-particle gaps will allow us to determine the boundaries of this phase and to compute the corresponding critical exponents.

For nonvanishing fields and setting $J=1/2$, the Hamiltonian can be written as
%
%
%%%%%%%%%%%%%%%%%%%%
\begin{equation}
 \label{eq:ham_WF}
 H=-N+Q+T_0+T_{+2}+T_{-2},
\end{equation}
%%%%%%%%%%%%%%%%%%%%
%
%
where now $Q$ counts the total number of charges and fluxes. The $T_n$ operators are linear in $h_x$ and $h_z$ and satisfy as previously $[Q,T_n]=n\, T_n$. Their precise expressions do not bring any special physical insight and are thus omitted here.

Once again, such a form is well suited to a PCUT treatment. We emphasize that our study  amounts to computing transition amplitudes of the effective QP-conserving Hamiltonian between the highly entangled eigenstates of the TCM \cite{Kitaev03}. To this end, it is essential to keep track of the anyon positions and of the underlying spin background simultaneously.
This makes our perturbation theory more complicated than the one derived in Ref.~\cite{Banks81} whose unperturbed Hamiltonian corresponds (in the gauge-theoretical reformulation used in
Ref.~\cite{Tupitsyn08}), to $J_p=h_z=0$, which has separable eigenstates.

We have obtained the ground-state energy per spin $e_0$, as well as the dispersions of the QP (dressed charges and fluxes) which are obviously mapped one onto the other when exchanging $h_x$ and $h_z$. A typical dispersion is displayed in Fig.~\ref{fig:Disp}. The gap $\Delta$ is the minimum of both dispersions, and we give it here in the region $h_x\leqslant h_z$ where charges are the lowest-energy excitations. 
Both $e_0$ and $\Delta$ were computed at order 8 in $(h_x,h_z)$, and we obtained
%
%
%%%%%%%%%%%%%%%%%%%%
\begin{eqnarray}
 \label{eq:gseSF}
 e_0&=& -\frac{1}{2} - \frac{1}{2}\big(h_z^2+h_x^2\big) -\frac{15}{8}\big(h_z^4+h_x^4\big)+\frac{h_x^2 h_z^2}{4} \nonumber \\
 && -\frac{147}{8}\big(h_z^6+h_x^6\big)+\frac{113}{32}\big(h_x^2 h_z^4+h_x^4 h_z^2 \big) +\frac{20869}{384} h_x^4 h_z^4 \nonumber \\
 &&  +\frac{6685}{128}\big(h_x^2 h_z^6+h_x^6 h_z^2 \big) - \frac{18003}{64}
 \big(h_z^8+h_x^8\big), \\
 \label{eq:gapSF}
 \Delta&=& 1-4 h_z-4 h_z^2-12 h_z^3+2 h_x^2 h_z- 36 h_z^4 + 3 h_x^2 h_z^2  \nonumber \\
 && +5 h_x^4-176 h_z^5+\frac{83}{4} h_x^2 h_z^3+\frac{27}{2} h_x^4 h_z-\frac{2625}{4} h_z^6  \nonumber \\
 && +63 h_x^2 h_z^4+71 h_x^4 h_z^2+92 h_x^6-\frac{14771}{4} h_z^7 \nonumber \\
 &&+\frac{28633}{64} h_x^2 h_z^5+\frac{925}{4} h_x^4 h_z^3+\frac{495}{2} h_x^6 h_z \nonumber \\
 &&- \frac{940739}{64} h_z^8+\frac{118029}{64} h_x^2 h_z^6+\frac{19263}{16} h_x^4 h_z^4 \nonumber \\
 &&+ \frac{80999}{96} h_x^6 h_z^2+\frac{495}{2} h_x^6 h_z^2+\frac{35649}{16} h_x^8.
\end{eqnarray}
%%%%%%%%%%%%%%%%%%%%
%
%

When one of the magnetic field components vanishes, the TCM is equivalent to the
Ising model in a transverse field so that setting $h_x=0$, we recover the
results obtained by He {\it et al.} \cite{He90}. Equation (\ref{eq:gapSF})
predicts a gap which vanishes continuously along a line starting from the
critical point $[h_x=0,h_z=0.1642(2)]$ and ending at the multicritical point
$[h_x=h_z=0.1703(2)]$.  As previously, error bars are obtained from various Dlog Pad\'e approximants used to analyze the series. 
%
%
%%%%%%%%%%%%%%
\begin{figure}[t]
 \includegraphics[width=0.75\columnwidth]{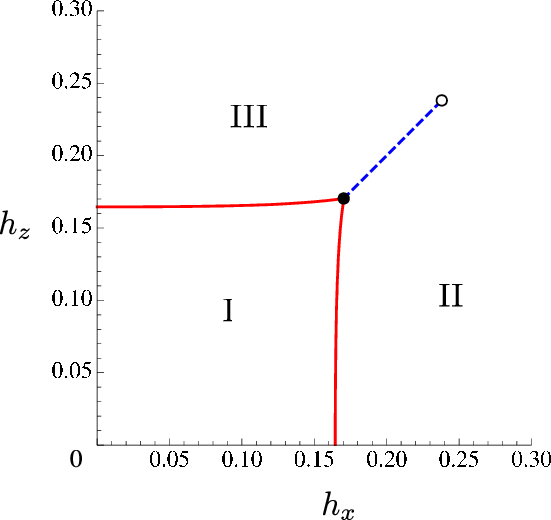}
 \caption{(Color online). Phase diagram in the plane $(h_x,h_z)$ for $J=1/2$,
   where second-order transition lines are drawn full (red), the
   first-order transition line is dashed (blue), the multicritical point
   is a full circle and the critical point is represented as an empty circle. Phases I, II and
   III are discussed in the text.}
 \label{fig:PD}
\end{figure}
%%%%%%%%%%%%%%
%
%
We also computed the critical exponent driving the closure of the gap. We found $\nu_{h_x=0}\simeq 0.65$, close to the expected Ising exponent and
$\nu_{h_x=h_z}\simeq 0.70$ at the multicritical point (note that at order 8, these
exponents are not yet fully converged). For intermediate values,
the exponent sticks to the Ising value except in the vicinity of the
multicritical point, indicating that phase transitions are Ising-like, except at
the multicritical point. This is confirmed by our QP picture. Away
from this point, only one kind of particle condenses (charge or flux), the other
one remaining gapped. At the multicritical point, both types of particles
condense simultaneously, and their mutual semionic statistics should become important and
gives the transition an unconventional character.

%
%
%%%%%%%%%%%%%%%%%%%%%%%
%%%%%%%%%%%%%%%%%%%%%%%
\section{Discussion}
%%%%%%%%%%%%%%%%%%%%%%%
%%%%%%%%%%%%%%%%%%%%%%%
%
%
The phase diagram obtained from our analytical calculations is shown in Fig.~\ref{fig:PD}.
Second order transition lines are obtained from the small-field expansion, the first-order line from the
classical (large-$S$) analysis and the position of the critical point from the
large-field expansion. As in Ref.~\cite{Tupitsyn08}, phase I is a
topological phase where $\langle A_s \rangle \simeq 1$ and
$\langle B_p \rangle \simeq 1$ in the ground state for $h_x,h_z\ll J$. This
phase has dispersive charge and flux excitations. Phase II (III) is such that
$\langle\sigma^x_i\rangle\simeq 1$ if $h_x\gg h_z\gg J$
($\langle\sigma^z_i\rangle\simeq 1$ if $h_z\gg h_x\gg J$), and
these phases have dispersive spin-flip excitations. 
However, using Eq.~(\ref{eq:gseLF}) and 
the Hellmann-Feynman theorem for the ground state energy one can compute $\langle\sigma^x_i\rangle$ and $\langle\sigma^z_i\rangle$ in phase I, and check that they do not vanish. 

Actually, no local order parameter can be used to characterize these various phases showing that the previous description is very rough. Furthermore, simple non-local order parameters can only be found on the Ising lines ({\it e.g.} if $h_x=0$, a semi-infinite
string of $\sigma^z$ operators on a line of the square lattice of
Fig.~\ref{fig:lattice}). Finding order parameters for these phases thus remains challenging. 

Apart from the phase diagram, the central result of our work is the set up of a QP picture for fluxes and charges in the topological phase. As a consequence, we have been able to compute the phase boundaries and the critical exponents by studying the locus of points where the 1-QP gap vanishes.
This QP description offers a wide range of perspectives. Indeed, the present PCUT approach 
is particularly well adapted to study the many-QP physics \cite{Knetter03_2}. 
This will allow us to investigate the likely existence of bound states in this model made up of flux-charge composites (fermionic statistics) that may change the critical properties.
Such fermions are expected to play a major role when switching on a
magnetic field in the $y$-direction since at lowest order, such a field induces a hopping of these fermions (as well as a local transmutation of two charges into two fluxes and vice-versa). 
Note that it may be difficult to study this problem using Monte Carlo simulations because of the usual sign problem, which arises in the presence of a transverse field. 

\acknowledgments

We wish to thank M. Kamfor, A. Kitaev, Z. Nussinov, and  G. S. Uhrig for fruitful and inspiring
discussions. K.P.S. acknowledges ESF and EuroHorcs for funding through his EURYI.

%\bibliography{/Users/Julien/Documents/Tex/Bibliotheques/bibliotheque2}

\begin{thebibliography}{26}
\bibitem[{Tem()}]{Tempo}
\bibinfo{note}{We confirmed this hypothesis by numerical studies.}
\end{thebibliography}

\end{document}